\documentclass[twocolumn]{autart}    

\usepackage{graphicx}          
\usepackage{amsmath,amssymb,amsfonts}
\usepackage{algorithmic}
\usepackage{graphicx}
\usepackage{textcomp}
\usepackage{xcolor}

\begin{document}

\begin{frontmatter}

\title{Synthesis of Functional Unknown Input Observers for LTV and MIMO LTI Systems with Arbitrary Relative Degree\thanksref{footnoteinfo}} 

\thanks[footnoteinfo]{Supported by the Ministry of Science and Higher Education of the Russian Federation (project no. FSER-2025-0002).}

\author[itmo]{Alexey A. Margun}\ead{alexeimargun@gmail.com},    
\author[itmo]{Alexey A. Bobtsov}\ead{bobtsov@mail.ru},               
\author[inria]{Denis V. Efimov}\ead{Denis.Efimov@inria.fr},  
\author[itmo]{Alexandr D. Panin}\ead{312636@niuitmo.ru}, 
\author[itmo]{Mariia O. Rassolova}\ead{rassolovam.01@gmail.com}

\address[itmo]{ITMO University, Saint Petersburg, Russia}  
\address[inria]{Inria, University Lille, CNRS, Lille, France}             


\begin{abstract}                          
This article focuses on the development of functional unknown input observers for systems with arbitrary relative degree. Two distinct approaches are presented to address this challenge. The first approach is tailored to a class of time-varying systems expressed in a canonical controllable form. This method leverages the Generalized Parameter Estimation-Based Observer framework. The article derives the conditions for applicability of this solution and outlines the limitations on the number of estimable state variables. The second approach targets multi-input multi-output systems. In contrast to existing methods, the proposed solution is applicable to systems with arbitrary relative degree, significantly broadening its scope of application. The theoretical results are validated through simulation studies, which demonstrate the effectiveness of the proposed observers. 
\end{abstract}

\end{frontmatter}

\section{Introduction}
The task of estimating a particular function of the state vector, without requiring its complete reconstruction, is a common problem in many practical applications. One key example is the design of output controllers derived from state feedback frameworks, where it is enough to reconstruct the control function directly. Another important application is found in output regulation control, where one output is directly measurable, but a different, unmeasured output must be controlled to achieve a desired set point \cite{kravaris}. In such cases, it is often unnecessary to estimate the entire state vector.

The theoretical foundations of linear functional observers were first established in \cite{chen} and subsequently expanded upon in \cite{darouach} and \cite{korovin}, among others. In \cite{darouach}, the authors present a streamlined approach to designing functional observers for linear time-invariant multivariable systems. These observers are characterized by an order equal to the dimension of the vector being estimated, and the work provides necessary and sufficient conditions for their existence and stability. In parallel, \cite{korovin} explores the design of minimum-order observers for linear state functionals, extending the framework to systems with vector outputs. The conditions for designing such observers are expressed through the concept of functional detectability. Furthermore, \cite{darouach2} formulates the necessary and sufficient conditions for the existence of an asymptotic functional observer in terms of matrix inequalities.

Despite these advancements, a significant limitation in much of the existing literature is the assumption that the inputs are fully measurable. This assumption restricts the practical applicability of these observers, particularly in scenarios involving disturbances or unmodeled dynamics, where the input may not be entirely known or accessible.

An effective approach to address this challenge is the use of Unknown Input Observers (UIOs). UIOs are especially useful when the input signal is heavily corrupted by disturbance or where direct measurement of the input is impractical due to cost constraints or technical limitations. This robustness also makes them suitable for systems with additive nonlinearities, parametric uncertainties, and unmodeled dynamics, which can be treated as additional unknown inputs. However, the design of UIOs is generally more complex compared to conventional observers, posing a challenge in their implementation.

Unknown input observers were first introduced by N. Kobayashi in 1982 \cite{uio_4}. This result was further developed in \cite{uio_1}, \cite{uio_2}, for discrete-time systems \cite{uio_3,uio_5}, some classes of nonlinear systems \cite{uio_3,uio_5}, and time-varying systems \cite{uio_7}.

Generally, constructing UIO is feasible primarily for systems where the relative degree between unknown inputs and the output signal is one. For plants with a higher relative degree, stringent conditions must be met to build an observer. For example, researches like \cite{ichala}, \cite{floquet} necessitate measuring the derivatives of the system’s output signal. In works such as \cite{coutinho}, \cite{wang}, the system’s original dynamic equation must be separable to isolate the observable components in the output. Several results are devoted to the synthesis of functional UIOs, but these solutions are limited by conditions related to the relative degree of the plant \cite{trinh}. 

Some recent studies have moved away from explicit dependence on the relative degree, allowing their methods to be applied to a broader class of systems. For instance, the work in \cite{reza} proposed a method for designing minimal unknown-input multi-functional observers for MIMO linear systems and established necessary and sufficient conditions for their existence. In \cite{polytopic}, the observer synthesis procedure was enhanced and extended to plants in polytopic form. The paper \cite{uio_sylvester} introduced an iterative approach to design a minimal-order unknown input functional observer for linear time-invariant systems. This method is particularly convenient for implementation, as it relies on solving a generalized Sylvester equation. However, challenges remain regarding the determination of the maximum number of linear combinations of observable state variables and the selection of the functional matrix, leaving these as open problems.

Filling these gaps, two approaches to design functional UIOs are presented in this work. The first method is applicable to linear time-varying (LTV) systems that can be transformed into a chain of integrators with an unknown input and a time-dependent output matrix. The core idea relies on the Generalized Parameter Estimation-Based Observer (GPEBO) framework. The paper formulates the conditions for the existence of such an observer and proposes a systematic approach for constructing the functional matrix. The second method targets linear time-invariant (LTI) MIMO systems with arbitrary relative degree. In contrast to the first approach, this method is applicable to a broader class of systems and provides the ability to control the convergence rate of the estimates. This flexibility makes it particularly suitable for practical applications where tuning the observer's performance is critical.

The key advantages of the proposed observers structure are as follows: 
\begin{itemize} 
    \item applicability to systems with arbitrary relative degree; 
    \item explicit determination of the maximum number of observable linear combinations of state vector variables; 
    \item a straightforward algorithm for computing the functional matrix; 
    \item capability to observe the full state vector for specific classes of nonlinear systems. \end{itemize}
The trade-off for these advantages is the limitation on the admissible linear functionals that can be reconstructed using the proposed approaches. For the first observer, the number of estimable state variables depends on the structure of the output matrix. In contrast, for the second observer, the number of available linear combinations of state variables that can be estimated is determined by the system's order and its relative degree.

The paper is organized as follows. Section 2 presents functional unknown input observer synthesis for LTV systems; Section 3 describes the functional observer synthesis for MIMO systems with unknown input for a number of special cases and generalizion for arbitrary relative degree; Section 4 contains an example of using the obtained result for nonlinear systems; the simulation results are presented in Section 5.

\section{Functional unknown input observer synthesis for a class of LTV systems}
This section focuses on the synthesis of functional observers for a class of linear time-varying (LTV) systems with unknown inputs.
\subsection{Problem statement}
Consider a class of LTV plants described by the equations:
\begin{equation}
\label{plant1}
    \left\{
    \begin{aligned}
        & \dot{x}(t)=Ax(t)+Bf(t), \\
        & y(t) = C(t)x(t),
    \end{aligned}
    \right.
\end{equation}
where
$$
A = \begin{bmatrix}
    0 &   & I_{n-1}\\
    0 & \hdots & 0
\end{bmatrix},
B = \begin{bmatrix}
    0\\
    \hdots\\
    0\\
    1
\end{bmatrix},
C^T(t) = \begin{bmatrix}
    c_1(t)\\
    c_2(t)\\
    \hdots\\
    c_n(t)
\end{bmatrix}
$$
are known matrices, $I_{n}\in \mathbb{R}^{n\times n}$ is an identity matrix, $y(t)\in \mathbb{R}$ is a measured output, $x(t)\in\mathbb{R}^n$ is an unmeasured state vector, $f(t)\in \mathbb{R}$ is an unknown input, that can include disturbances,  parametric uncertainties and unaccounted dynamics. It should be noted that every observable and controllable linear SISO system can be transformed to form (\ref{plant1}).

The objective is to develop a functional unknown input observer that can estimate the part of the state vector:
$$
\lim_{t\xrightarrow{}\infty} |\bar{x}(t)-\hat{\bar{x}}(t)|=0,
$$
where $\bar{x}(t)=Qx(t)$, $\hat{\bar{x}}(t)$ is an estimate of the $\bar{x}$, $Q$ is a design matrix that meets the constraints outlined below.
\subsection{Functional unknown input observer synthesis}
Taking into account that the system under consideration (\ref{plant1}) is a chain of integrators, we express its output as:
\begin{equation}
\label{y1}
\begin{aligned}
    &y = c_1(t)x_1(t) + c_2(t)x_2(t)+\hdots+c_n(t)x_n(t)=\\
    &c_1(t)x_1(t) + c_2(t)\dot{x}_1(t)+\hdots + c_n(t)x_1^{(n-1)}(t).
\end{aligned}    
\end{equation}
Let $\beta$ denote the index of the last nonzero element of $C(t)$. For example, if $C\in \mathbb{R}^5$ and $\beta=3$ then
$$
C(t) = \begin{bmatrix}
    c_1(t) & c_2(t) & c_3(t) & 0 & 0
\end{bmatrix}.
$$
Let us express the highest derivative of the state vector that directly influences the output:
\begin{equation}
\label{high_der}
    x_1^{(\beta-1)} = \frac{1}{c_{\beta}}\left(-c_1x_1-\hdots-c_{\beta-1}x_1^{(\beta-2)}+y\right),
\end{equation}
and rewrite equation (\ref{high_der}) in state-space representation
\begin{equation}
\label{plant_q}
    \dot{w}(t)=R(t)w(t)+D(t)y(t),
\end{equation}
where
$$
R(t)=\begin{bmatrix}
    0 & & I_{\beta-2}\\
    -\frac{c_1(t)}{c_\beta(t)} & \hdots & -\frac{c_{\beta-1}(t)}{c_\beta(t)}
\end{bmatrix},
D(t)=\begin{bmatrix}
    0\\
    \hdots\\
    0\\
    1/c_\beta(t)
\end{bmatrix},
$$
$w(t)\in\mathbb{R}^{\beta-1}$ is a state vector. 

To solve the initial problem, it is necessary to estimate the vector $w(t)$.  For this purpose, we apply the generalized parameter estimation-based observer \cite{gpebo}. Introduce an observer as “copy” of the (\ref{plant_q}) using another dynamical system
\begin{equation}
\label{obs_gpebo}
    \left\{ \begin{aligned}
        &\dot{\xi}(t)=R(t)\xi(t)+D(t)y(t),\\
        &\dot{\Phi}(t)=R(t)\Phi(t), \Phi(0)=I_{\beta-1},
    \end{aligned} \right.
\end{equation}
where $\xi(t)$ is an observer state vector, $\Phi$ is a fundamental matrix.

Consider the dynamics of observation error:
\begin{equation}
\label{obs_error_1}
    e(t)=w(t)-\xi(t) \xrightarrow{} \dot{e}(t)=R(t)e(t).
\end{equation}
Taking into account (\ref{plant_q}) and (\ref{obs_gpebo}) the solution of (\ref{obs_error_1}) can be found as
\begin{equation}
\label{obs_error_solution}
    e(t)=\Phi(t)e(0).
\end{equation}
Therefore
\begin{equation}
    w(t)=\xi(t)+\Phi(t)e(0).
\end{equation}
From (\ref{obs_gpebo}) and (\ref{obs_error_solution}) it follows that if the matrix $R(t)$ is stable, then $e(t)$ asymptotically tends to zero and
$$
\lim_{t\xrightarrow{}\infty}\xi(t) = w(t),
$$
Next, we can use $\xi(t)$, (\ref{high_der}) and (\ref{plant_q}) to obtain estimates of $x_1(t)$ and reconstruct its $\beta-2$ derivatives.

Let us formulate the following theorem.
\begin{thm}
Let the vector $C(t)$ of the plant (\ref{plant1}) be such that the matrix $R(t)$ is stable. Then, the observer (\ref{obs_gpebo}) provides an estimate of $\bar{x}(t)=Qx(t)$ with asymptotic convergence and
$$
Q=\begin{bmatrix}
    I_{\beta-1} & 0_{(\beta-1)\times (n-\beta+1)}
\end{bmatrix}.
$$
\end{thm}
The proof follows from the above calculations.

\textbf{Remark 1} The proposed solution has a simple structure but does not allow for adjusting the convergence rate and is only applicable to systems with a stable matrix $R(t)$. In the next section, an alternative observer synthesis approach will be proposed to overcome these shortcomings.

\section{Functional unknown input observer synthesis for MIMO systems with arbitrary relative degree}
In this section, we consider the synthesis of a functional unknown input observer for the MIMO systems with an arbitrary relative degree.
\subsection{Problem statement}
Consider a linear MIMO time-invariant plant \cite{our_UIO_diag}:
\begin{equation} \label{plant}
    \left\{ 
    \begin{aligned} 
        &\dot{x}(t)=Ax(t)+Bf(t),\\ 
        &y(t)=Cx(t), 
    \end{aligned} \right.
\end{equation}
where $x(t)\in \mathbb{R}^n$ is the state vector, $f(t)\in\mathbb{R}^{m}$ is the unmeasured input signal vector, $y(t)\in\mathbb{R}^l$ is the measurable output signal vector, $A, B, C$ are known constant matrices of appropriate dimensions. The plant is characterized by a vector of relative degrees between the inputs and outputs $r=[r_1, \dots, r_l]$. This implies that the following condition is satisfied for $i=\overline{1, l}$:
\[
\begin{aligned}
    &c_ib_k = c_iA^jb_k = 0, \, \forall j = 1, \dots, r-2, \forall k=1,\dots,m,\, \textrm{and}\\
    &c_iA^{r-1}b_k\ne0, \exists k\in\{1,\dots,m\},
\end{aligned}
\]
where $c_i$ and $b_k$ are $i$-th row and $k$-th column of the matrices $C$ and $B$, respectively. It is worth highlighting that this condition is a fundamental characteristic of the dynamic system itself, meaning that every system possesses a relative degree \( r_i > 0 \). Consequently, it does not impose any limitations on the applicability of the method introduced in this paper.

The objective is to develop an unknown input functional observer that can estimate a specific portion of the state vector \( x \) or certain linear combinations of its variables. This desired part of the state vector is represented as:  
\[
\bar{x} = Qx \in \mathbb{R}^q,
\]  
where \( Q \) is an appropriately chosen matrix that meets the constraints outlined below.

\subsection{Unknown input observer synthesis for plants with unit relative degree}
In this section, we introduce an algorithm for synthesizing an unknown input observer. 

Define the following additional matrices:
\[
P = \begin{bmatrix}
C_1A^{r_1} \\
C_2A^{r_2} \\
\dots\\
C_l A^{r_l}
\end{bmatrix},
\]
\[
N = \begin{bmatrix}
    C_1A^{r_1-1}B_1 & C_1A^{r_1-1}B_2 & \dots & C_1A^{r_1-1}B_m \\
    C_2A^{r_2-1}B_1 & C_2A^{r_2-1}B_2 & \dots & C_2A^{r_2-1}B_m \\
    \vdots & \vdots & \ddots & \vdots \\
    C_lA^{r_l-1}B_1 & C_lA^{r_l-1}B_2 & \dots & C_lA^{r_l-1}B_m
\end{bmatrix},
\]
where we assume that the matrix $N$ has full column rank.

The observer is defined as follows \cite{our_UIO}:
\begin{equation}
	\dot{\hat{x}} = M\hat{x} + L(y - C\hat{x}) + Gy^{(r)} = F\hat{x} + Ly + Gy^{(r)},
	\label{observer}
\end{equation}
where \( y^{(r)}(t) = \begin{matrix}[y_1^{(r_1)} & y_2^{(r_2)} & \dots & y_l^{(r_l)}]\end{matrix} \), and \( y_i^{(j)} \) denotes the \( j \)-th derivative of the \( i \)-th system output. The matrix \( F = M - LC \), and \( M \) and \( G \) are determined to satisfy the following conditions:
\begin{equation}
\label{obs_eqn}
\begin{cases}
    B - GN = 0, \\
    M = A - GP.
\end{cases}
\end{equation}
Let us define the observation error as \( \tilde{x}(t) = x(t) - \hat{x}(t) \). Differentiating this expression and using \eqref{obs_eqn}, we derive the following dynamic model for the error:
\begin{align}
    \label{condition}
    \begin{aligned}
        &\dot{\tilde{x}}(t) = Ax(t) + Bf(t) - M\hat{x}(t) - L(y(t) - \\
        &C\hat{x}(t)) - Gy^{(r)} (t)= Ax(t) + Bf(t) - M\hat{x}(t) + \\
        &LC\hat{x}(t) - GPx(t) - GNf(t) - Ly(t) = \\
        &(A - GP)x(t) - M\hat{x}(t) - LC\tilde{x}(t) = \\
        &(M - LC)\tilde{x}(t) = F\tilde{x}(t).
    \end{aligned}
\end{align}
This shows that the dynamics of the observation error \( \tilde{x}(t) \) depend entirely on the matrix \( F \), which is determined during the observer design. By choosing \( L \) appropriately, the matrix \( F \) can be configured to ensure stability and exponential convergence of \( \tilde{x}(t) \) to zero.

\textbf{Remark 2}  
The system of equations in \eqref{condition} admits a unique solution for \( G \), expressed as:  \[
G = B(N^{\textrm{T}}N)^{-1}N^{\textrm{T}},
\]  
provided the following conditions are satisfied \cite{our_UIO_diag}:  
\begin{itemize}
    \item \( \textrm{rank}(N) = \textrm{rank}(B) \);  
    \item the matrix pair \( (M, C) \) is observable. 
\end{itemize}
 However, the observer described in \eqref{observer} cannot be implemented directly because the derivatives of the output \( y \) are typically not available in practice.  Without loss of generality, let us assume that $r_1 \le r_2 \le \hdots \le r_l$. To allow the construction of an observer for a system with relative degrees $r_i = 1, i = \overline{1, l}$, we introduce $r_l$ auxiliary variables as follows:
\begin{align*}
    \begin{aligned}
    z_1(t)&=\hat{x}(t) - Gy^{(r-1)}(t), \\
    \dot{z}_1(t)&=F(z_1(t)+Gy^{(r-1)}(t)) + Ly(t),\\
    z_2(t)&=z_1(t)-FGy^{(r-2)}(t), \\
    \dot{z}_2(t)&=F(z_2(t)+FGy^{(r-2)}(t)) + Ly(t), \\
    z_3(t)&=z_2(t)-F^2Gy^{(r-3)}(t),\\
    &\hdots\\
    \dot{z}_{r_1}(t)&=Fz_{r_1}(t)+F^{r_1}G
        \left[
        \begin{matrix}
            y_1(t)\\
            0\\
            \hdots\\
            0
        \end{matrix}
        \right]+F^{r_1}G
        \left[
        \begin{matrix}
            0\\
            y_2^{(r_2-r_1)}(t)\\
            \hdots\\
            y_l^{(r_l-r_1)}(t)
        \end{matrix}
        \right] +\\
    Ly(t), \\
    \end{aligned}
\end{align*}
\begin{align*}
    \begin{aligned}
    &\hdots\\
    \dot{z}_{r_l}(t)&=Fz_{r_l}(t)+F^{r_1}G
    \left[
    \begin{matrix}
        y_1(t)\\
        0\\
        \hdots\\
        0
    \end{matrix}
    \right]+\hdots + F^{r_2}G
        \left[
        \begin{matrix}
            0\\
            y_2(t)\\
            \hdots\\
            0
        \end{matrix}
        \right]+\\
    &\hdots +F^{r_l}G
    \left[
    \begin{matrix}
        0\\
        0\\
        \hdots\\
        y_l(t)
    \end{matrix}
    \right]+ Ly(t).
    \end{aligned}
\end{align*}
Using reference variables, an estimate of the state vector can be obtained as follows
\begin{align}
\label{uio_aux}
    \begin{aligned}
    &\hat {x}=z_1 + Gy^{(r-1)}=z_{r_l}+F^{r_l-1}G
        \left[
        \begin{matrix}
            0\\
            0\\
            \hdots\\
            y_l 
        \end{matrix}
        \right]+ \hdots + \\
    &F^{r_1}G
        \left[
        \begin{matrix}
            0\\
            y_2^{(r_2-r_1-1)}\\
            \hdots\\
            y_\beta^{(r_\beta-r_1-1)} 
        \end{matrix}
        \right] + \hdots + G
        \left[
        \begin{matrix}
            y_1^{(r_1 - 1)}\\
            y_2^{(r_2-1)}\\
            \hdots\\
            y_\beta^{(r_l-1)} 
        \end{matrix}
        \right],\\
    \end{aligned}
\end{align}
\begin{align}
    \begin{aligned}
    &\dot{z}_{r_l}(t)=Fz_{r_l}(t)+F^{r_1}G
    \left[
    \begin{matrix}
        y_1(t)\\
        0\\
        \hdots\\
        0
    \end{matrix}
    \right]+\hdots + F^{r_2}G
        \left[
        \begin{matrix}
            0\\
            y_2(t)\\
            \hdots\\
            0
        \end{matrix}
        \right]+\\
    &\hdots +F^{r_l}G
    \left[
    \begin{matrix}
        0\\
        0\\
        \hdots\\
        y_l(t)
    \end{matrix}
    \right]+ Ly(t).
    \end{aligned}
\end{align}
For the sake of brevity, we shall reformulate the observer as follows:
\begin{equation}
\label{uio_brev}
\left\{
\begin{matrix}
    \hat{x}(t)=W_1(z_{r_l}(t), y^{(r_l - 1)}(t)),\\
    \dot{z}_{r_l}= W_2(z_{r_l}(t), y(t)).
\end{matrix}
\right.
\end{equation}
This state observer is applicable to systems with a unit relative degree, i.e., if $r_i=1, i=\overline{1,l}$, then $\hat{x}$ can be constructed using the information about $y$ only. Moreover, it can be used as a foundation for constructing a functional observer for systems with arbitrary relative degree. The method for constructing such observers will be presented in the next section.

\subsection{Functional unknown input observer synthesis for plants with arbitrary  relative degree}
In this section we will first exemplify the synthesis procedure for systems with different relative degrees, and next we will present a generalized method for systems with an arbitrary relative degree.

To this end, we will equip the observer (\ref{uio_brev}) with the output matrix $Q$, which allows us to estimate the vector $\bar{x}$ by
\begin{equation}
\hat{\bar x} = Q \hat{x}
\label{func_bar_x}
\end{equation}
and the matrix $Q$ needs to be chosen in a way providing possibility to implement the observer.

First, consider a system with two outputs and a vector of relative degrees $r=[1, 2]$. The state estimate can be expressed as:
\begin{equation}
    \hat{\bar{x}}=Q\hat{x}=Qz_2+QFG
    \begin{bmatrix}
        0\\
        y_2
    \end{bmatrix}+QG
    \begin{bmatrix}
        y_1\\
        \dot{y}_2
    \end{bmatrix}.
\end{equation}
To synthesize a functional observer, it is necessary that \( QG = 0 \). Since \( G \) is linearly dependent on \( B \), the condition can be modified to $QB=0$. In such a case the expression for \( \bar{x} \) simplifies to:
$$
\hat{\bar{x}} = Q\left(z_2 + FG\begin{bmatrix}
        0\\
        y_2
    \end{bmatrix}\right).
$$
The resulting expression does not contain immeasurable signals, which means that the observer can be constructed.

Next, consider a system with three outputs and a vector of relative degrees $r=[1, 2, 3]$. The state estimate can be expressed as:
\begin{equation}
    \hat{\bar{x}}=Q\hat{x}=Q\left(z_3+F^2G
    \begin{bmatrix}
        0\\
        0\\
        y_3
    \end{bmatrix}+FG
    \begin{bmatrix}
        0\\
        y_2\\
        \dot{y}_3
    \end{bmatrix}
    +G
    \begin{bmatrix}
        y_1\\
        \dot{y}_2\\
        \ddot{y}_3
    \end{bmatrix}\right).
\end{equation}
The state $ \bar{x} $ can be observed if $QG=0$ and $ QFG = 0 $. Taking into account, that $CB=0$ and linear dependence of $B$ and $G$, we obtain
$$
\begin{aligned}
    QFG &= Q(M - LC)G = Q(A - GCA^3 - LC)G \\
    &= QAG - QGCA^3G - QLCG \\
    &= QAG.
\end{aligned}
$$
Thus, $ \bar{x} $ can be observed if $QG=0$ and $ QAG = 0 $, or equivalently, if $QB=0$ and $QAB=0$.

Similarly, we obtain an estimate $\hat{\bar{x}}$ for a plant with three outputs and a vector of relative degrees $r=[1, 2, 4]$:
\begin{equation}
\begin{aligned}
    &\hat{\bar{x}}=Q\left(z_4+F^3G
    \begin{bmatrix}
        0\\
        0\\
        y_3
    \end{bmatrix}+F^2G
    \begin{bmatrix}
        0\\
        0\\
        \dot{y}_3
    \end{bmatrix}
    +\right.\\
    &\left. FG
    \begin{bmatrix}
        0\\
        y_2\\
        \ddot{y}_3
    \end{bmatrix}+G
    \begin{bmatrix}
        0\\
        \dot{y}_2\\
        y^{(3)}_3
    \end{bmatrix}
    \right).
\end{aligned}
\end{equation}
The state $ \bar{x} $ can be observed if $QG=0$, $QFG=0$ (according to the previously given computations, this property is verified if $QAG=0$) and $QF^2G = 0 $, and 
$$
\begin{aligned}
    QF^2G &= Q(M - LC)^2G \\
    &= QM^2G - QLCMG - QMLCG + QLCLCG \\
    &= Q(A - GCA^4)^2G - QLC(A - GCA^4)G \\
    &= (QA^2G - QGCA^5G - QAGCA^4G\\
    & +QGCA^4GCA^4G) \\
    &\quad - QLCAG + QLCGCA^4G \\
    &= QA^2G.
\end{aligned}
$$
Thus, \( \bar{x} \) can be observed if: $ QG=QAG =QA^2G = 0 $, i.e., $ QB=QAB = QA^2B = 0 $.

Generalizing the obtained results, let us formulate the following theorem.
\begin{thm}
Let the pair $(A,C)$ be detectable and the matrix \( Q \) satisfy the condition:
\begin{equation}
\label{theorem_condition}
QA^iB = 0, \quad i = 0, \dots, r_l-2,
\end{equation}
then there is a gain $L$ such that the observer (\ref{uio_brev}), (\ref{func_bar_x}) exponentially estimates for the system (\ref{plant}) the variable $\bar{x}$ that belongs to the unmeasured subspace of dimension $q=n-r_l$.
\end{thm}
\textbf{Proof of the Theorem 2.} 
The maximal number of observable linear combinations of the state variables depends on the rank of the matrix \( Q \in\mathbb{R}^{q\times n} \), and in the best case \( q = \operatorname{rank}(Q) \). To determine the restrictions on $q$, consider the subspace:
\[
\nu = \operatorname{span}\{B, AB, \dots, A^{r_l-2}B\}.
\]
The dimension of this subspace is given by:
\[
\rho = \operatorname{rank}(\nu) \leq r_l - 1,
\]
because the basis of \( \nu \) can contain at most \( r_l-1 \) linearly independent vectors, based on the relative degree \( r_l \) of the system.
To satisfy the condition in \eqref{theorem_condition}, the rows of \( Q \) must belong to the orthogonal complement of \( \nu \), denoted as \( \nu^* \). The dimension of \( \nu^* \) is
\[
\operatorname{dim}(\nu^*) = n - \rho.
\]
Thus, the rank of \( Q \) is bounded by:
\[
\operatorname{rank}(Q) \leq \operatorname{dim}(\nu^*) = n - \rho = n - (r_l - 1) = n - r_l + 1.
\]
This shows that the observer can estimate up to \( q = n - r_l + 1 \) distinct linear combinations of the state vector variables.  
Note that the output variable belongs to this subspace by definition of the relative degree $r_l$. Since $y_l$ is available for measurement, it can be always extracted from $\nu^*$, and finally the dimension of the unmeasured subspace that can be reponstructed by (\ref{uio_brev}), (\ref{func_bar_x}) is $n-r_l$.

\textbf{Remark 3.} There is a straightforward method to calculate \( Q \):
\begin{enumerate}
    \item The rows of \( Q \) must lie in the orthogonal complement of the subspace spanned by the vectors: 
    $$
    T=[B, AB, .. A^{r-2}B].
    $$
    \item Determine a basis for the kernel of \( T \), for instance, using the Singular Value Decomposition (SVD) or the Gram-Schmidt orthogonalization method.
    \item Form the matrix \( Q \) by using the rows of this kernel basis.
\end{enumerate}
\textbf{Example.} Let's consider the calculation of the matrix for the following system with relative degree $r=3$
$$
A=\begin{bmatrix}
    0 & 1 & 0 & 0\\
    0 & 0 & 1 & 0\\
    0 & 0 & 0 & 1\\
    0 & 0 & 0 & 0
\end{bmatrix}, B=
    \begin{bmatrix}
        0 \\
        0 \\
        0 \\
        1
    \end{bmatrix},
    C^T=\begin{bmatrix}
        1 \\
        1 \\
        0 \\
        0
    \end{bmatrix}.
$$
Calculate \( T = [B, AB] \):
$$
T=[B, AB]=\begin{bmatrix}
    0 & 0\\
    0 & 0\\
    0 & 1\\
    1 & 0
\end{bmatrix}.
$$
Find a basis for \( \ker(T) \):
$$
\textrm{ker}\{T\}=\begin{bmatrix}
    1 & 0 & 0 & 0\\
    0 & 1 & 0 & 0
\end{bmatrix}.
$$
Clearly the sum of these two rows gives the output matrix, hence, one of the rows can be utilized to define $Q$, or we can just select the orthogonal to $C$ vector in $\textrm{ker}\{T\}$: $Q = \begin{bmatrix} 1 & -1 & 0 & 0 \end{bmatrix} $.

\section{Unknown input observer synthesis for a class of nonlinear systems}
In some cases, the proposed method can be adapted to estimate the state variables of systems with additive nonlinearities. To illustrate this, consider a bilinear BIBO plant:
\begin{equation}
	\left\{
		\begin{aligned}
		&\dot{x} = Ax + B\Phi(x),\\
		&y = Cx,
		\end{aligned}
	\right.
\end{equation}
where $A, B, C$ are taken from previous example,
$$
\Phi(x) = x_1x_3 = x_1(\dot{y}-x_2).
$$
Using the proposed algorithm, it is possible to estimate the variables $\hat{x}_1$ and $\hat{x}_2$ treating $\Phi$ as unknown input. 
Define the full state vector estimate by $\zeta$ and construct an observer:
\begin{equation}
    \begin{aligned}
        \dot{\zeta}&=A\zeta + K(y-C\zeta)+B\Phi(\hat{x})\\
        &=A\zeta +K(y-C\zeta) + B\hat{x}_1\dot{y}-B\hat{x}_1\hat{x}_2
    \end{aligned}
\label{observer_zeta}
\end{equation}
where $K$ is designed such that $A-KC$ is Hurwitz.
The error dynamics are given by:
\begin{equation}
\begin{aligned}
    &\dot{\tilde{\zeta}} = \dot{x} - \dot{\zeta}=Ax+Bx_1\dot{y}-Bx_1x_2-A\zeta-\\
    &K(y-C\zeta)-B(x_1-\tilde{x}_1)\dot{y}+B(x_1-\tilde{x}_1)(x_2-\tilde{x}_2)=\\
    &(A-KC)\tilde{\zeta}+B\dot{y}\tilde{x}_1-B(\tilde{x}_1x_2+x_1\tilde{x}_2-\tilde{x}_1\tilde{x}_2),
\end{aligned}    
\end{equation}
where $\tilde{x}_i = x_i-\hat{x}_i$ for $i=1,2$ is the exponentially converging error of the functional UIO, and $\tilde\zeta = x-\zeta$ is the estimation error for the observer (\ref{observer_zeta}).
Since the considered system is BIBO and the estimates of the first two state variables exponentially converge to zero, the dynamics of the full-state observation error is asymptotically reduced to the stable system:
$$
\dot{\tilde{\zeta}} = (A-KC)\tilde{\zeta},
$$
which guarantees asymptotic estimation of the full state vector $x$.
The implementation of the observer (\ref{observer_zeta}) requires $\dot{y}$. To overcome this limitation introduce additional variable
\begin{equation}
    \xi = \zeta - B\hat{x}_1y
    \label{xi}
\end{equation}
and consider its derivative
\begin{equation}
\begin{aligned}
    \dot{\xi}&=\dot{\zeta}-B\dot{\hat{x}}_1y-B\hat{x}_1\dot{y}\\
    &=A\zeta+K(y-C\zeta)-B\dot{\hat{x}}_1y-B\hat{x}_1\hat{x}_2 \\
    &=A(\xi+B\hat{x}_1y)+Ky-KC(\xi+B\hat{x}_1y)-B\dot{\hat{x}}_1y-\\
    &B\hat{x}_1\hat{x}_2.
\end{aligned}
\label{observer_xi}
\end{equation}
To compute $\dot{\hat{x}}_1$, use the functional UIO from Theorem 1:
\begin{equation}
    \left\{
    \begin{aligned}
        &\hat{\bar{x}}=Qz_2+QFGy,\\
        &\dot{z}_2=Fz_2+F^2Gy+Ly,
    \end{aligned}
    \right.
\end{equation}
where $\hat{\bar{x}}^T=[\hat{x}_1 \, \hat{x}_2]$, $z_2^T = [z_{21} \, z_{22} \, z_{23} \, z_{24}]$. Considering $QFG=Q(A-GCA^2-LC)G=0$, it follows that:
\begin{equation}
    \dot{\hat{x}}_1=\dot{z}_{21}=z_{22}.
    \label{hat_x1}
\end{equation}
Thus, observer (\ref{observer_xi}) is implementable, and using (\ref{xi}) allows to estimate full state vector of the plant.

\section{Simulation results}
To test the performance of the proposed approach, we will conduct computer simulation of both observer synthesis methods.
\subsection{Simulation of the observer for LTV system}
Consider a SISO LTV plant in state-space form described by matrices
$$
A=\begin{bmatrix}
    0 & 1 & 0 & 0\\
    0 & 0 & 1 & 0\\
    0 & 0 & 0 & 1\\
    -1 & -4 & -6 & -4
\end{bmatrix},
B=\begin{bmatrix}
    0\\
    0\\
    0\\
    1
\end{bmatrix},
C^T=\begin{bmatrix}
    1\\
    2+\sin 0.3t\\
    1\\
    0
\end{bmatrix},
$$
with initial conditions $x^T(0)=\begin{bmatrix}
    1 & 0.5 & -0.5 & 0
\end{bmatrix}$

\begin{figure}[h!]
    \centering
    \includegraphics[width=1\linewidth]{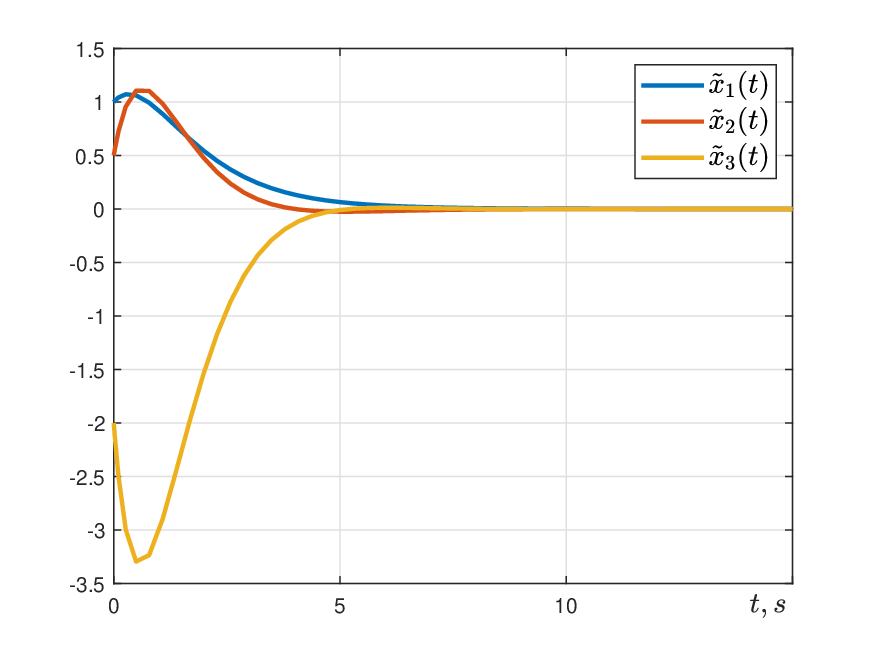}
    \caption{Observation errors of the functional observer for SISO system}
    \label{fig:SISO}
\end{figure}

Rewrite it in the form (\ref{plant1}):
$$
\dot{x}(t)=A_0x(t)+Bf(t),
$$
where
$$
A_0=\begin{bmatrix}
    0 & I_{3}\\
    0 & 0
\end{bmatrix},
f(t)=u(t)+\begin{bmatrix}
      -1 & -4 & -6 & -4
\end{bmatrix}x(t).
$$
Observer matrices are calculated as follows
$$
Q(t)=\begin{bmatrix}
    -2-\sin 0.3t & 1\\
    -1 & 0
\end{bmatrix},
D = \begin{bmatrix}
    0\\
    1
\end{bmatrix}.
$$
Transients of observation error for $x_1(t)$, $x_2(t)$ and $x_3(t)$ are shown in Figure \ref{fig:SISO}. It can be seen that the estimate of the state variable asymptotically converges to the true value.

\subsection{Simulation of the observer for MIMO system}
Consider the following plant:
$$
    A =
    \begin{bmatrix}
        0 & 1 & 0 & 0 & 0 \\
        0 & 0 & 1 & 0 & 0 \\
        0 & 0 & 0 & 1 & 0 \\
        0 & 0 & 0 & 1 & 1 \\
       -1 & -2 & -3 & -5 & -5 \\
    \end{bmatrix}, 
    B =
    \begin{bmatrix}
        0 \\
        0 \\
        0 \\
        0 \\
        1 \\
    \end{bmatrix}, 
    C^T =
    \begin{bmatrix}
        1 & 1 \\
        1 & 0 \\
        0 & 1 \\
        0 & 0 \\
        0 & 0 \\
    \end{bmatrix},
$$
where the relative degree of the system is \( r = 3 \).

To design the observer, we calculate the matrices \( G \), \( M \), and \( L \) such that the eigenvalues of the closed-loop system are \( \{-4, -5, -6, -7, -8\} \). The resulting observer matrices are:
$$
    G =
    \begin{bmatrix}
        0 & 0 \\
        0 & 0 \\
        0 & 0 \\
        0 & 0 \\
        0 & 1 \\
    \end{bmatrix}, 
    M =
    \begin{bmatrix}
        0 & 1 & 0 & 0 & 0 \\
        0 & 0 & 1 & 0 & 0 \\
        0 & 0 & 0 & 1 & 0 \\
        0 & 0 & 0 & 0 & 1 \\
        0 & 0 & 0 & -1 & 0 \\
    \end{bmatrix}, 
    L =
    \begin{bmatrix}
        -11.20 &  0.30 \\
        23.28  &  0.58 \\
        7.42   & 17.62 \\
       -66.96  & 102.93 \\
      -129.54  & 174.26 \\
    \end{bmatrix}.
$$
Next, we construct the subspace:
\[
\nu = \operatorname{span}\{B, AB, \dots, A^{r-2}B\} =
\begin{bmatrix}
    0 & 0 \\
    0 & 0 \\
    0 & 0 \\
    0 & 1 \\
    1 & -5 \\
\end{bmatrix}.
\]
The rows of \( Q \) are chosen to be in the orthogonal complement of \( \nu \):
\[
Q =
\begin{bmatrix}
    1 & 0 & 0 & 0 & 0 \\
    0 & 1 & 0 & 0 & 0 \\
    0 & 0 & 1 & 0 & 0 \\
\end{bmatrix}.
\]
Thus, the proposed functional observer allows the estimation of the first three state variables.
The initial conditions are chosen as:
\[
x(0) = \begin{bmatrix} 1 & -1 & 0.3 & -0.5 & 0 \end{bmatrix}^T.
\]
\begin{figure}[h!]
    \centering
    \includegraphics[width=1\linewidth]{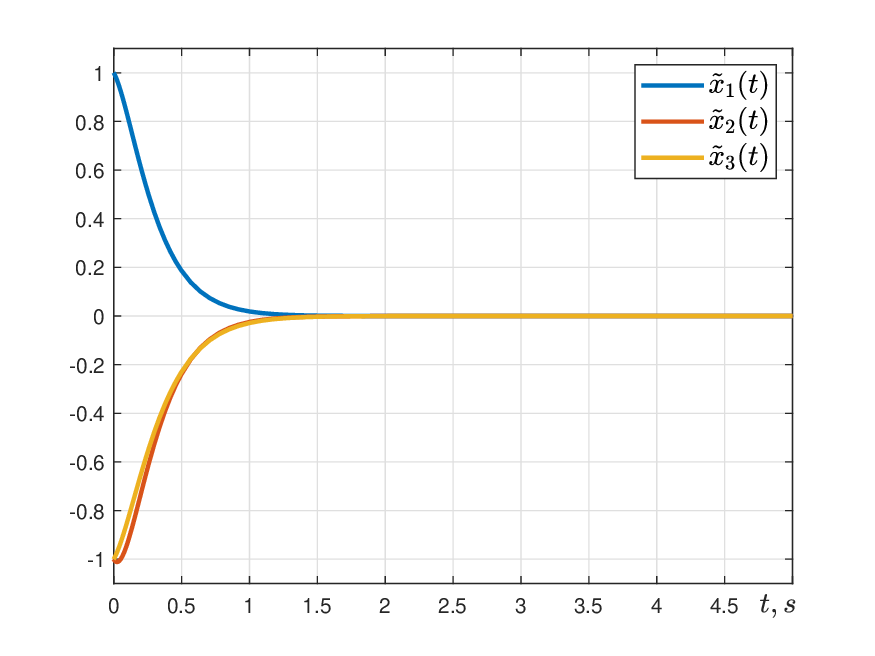}
    \caption{Observation errors of the functional observer for MIMO system.}
    \label{fig:MIMO}
\end{figure}
Figure \ref{fig:MIMO} shows that the estimation errors of the three state variables converge to zero, which confirms the observer's performance.

\section{Conclusion}
The article proposes two new methods for constructing functional state vector observers under conditions of unmeasured input. The first approach is applicable to systems that can be represented as a chain of integrators with an unknown input signal and a LTV output matrix. An algorithm for constructing a functional matrix and a condition for observer stability based on the output matrix are presented. The second approach allows synthesizing functional unknown input observers for MIMO systems. An algorithm for calculating a functional matrix is proposed and a proof of the stability of this approach is presented. Computer modeling confirms the efficiency of both approaches.





\bibliographystyle{plain}        
\bibliography{autosam}           




\end{document}